\documentclass[letterpaper,twocolumn,10pt]{article}
\usepackage{usenix}

\usepackage{amsmath}
\usepackage{graphicx}
\usepackage{hyperref}
\usepackage{pgf}
\usepackage{longtable}
\usepackage{setspace}
\usepackage{moreverb}
\usepackage[justification=centering]{caption}

\date{}
\title{Security impact ratings considered harmful}
\author{\rm{Jeff Arnold, Tim Abbott, Waseem Daher, Gregory Price,}\\
\rm{Nelson Elhage, Geoffrey Thomas, Anders Kaseorg}\\
Massachusetts Institute of Technology}

\begin{document}

\maketitle

\begin{abstract}
In this paper, we question the common practice of assigning security impact
ratings to OS updates.  Specifically, we present evidence that ranking updates
by their perceived security importance, in order to defer applying some
updates, exposes systems to significant risk.

We argue that OS vendors and security groups should not focus on security
updates to the detriment of other updates, but should instead seek update
technologies that make it feasible to distribute updates for all disclosed OS
bugs in a timely manner.

\end{abstract}

\section{Introduction}

Today, OS vendors and other computer security groups track and publish security
impact information to ``provide a simple way to judge the severity of security
updates''~\cite{security-response-team}.  OS vendors use this information
internally in order to determine which updates should be sent to customers in a
timely manner.  System administrators rely on this information ``to better
schedule upgrades to their systems''~\cite{security-response-team}---in other
words, to decide whether an upgrade needs to happen immediately or whether it
can be delayed for weeks or even months until the next ``important'' upgrade
comes along.

In this paper, we argue that this general approach to OS
security---specifically, tracking security updates separately from other
bug-fix updates so that security updates can be applied long before the average
update---is counter-productive to OS security.

We show that the security implications of OS bugs can easily elude developers,
so that the true security implications of bugs are commonly not discovered until
weeks or months after the bugs have been publicly disclosed.  During this
period, the patch for correcting a bug can remain widely unused since the bug
has no known security impact.  We present evidence that finding dangerous
high-impact attacks for these disclosed ``low-impact'' bugs is much easier than
finding new previously-unknown problems.  Every disclosed\footnote{Since
exploits can be generated from binary updates alone~\cite{Brumley}, bug
disclosure has occurred even if only a binary update has been published.} bug
that is classified as having low impact is therefore potentially an invaluable
blueprint for attackers to achieve their goals.

Tracking, classifying, and prioritizing security updates to the detriment of
other updates is therefore a major security liability for operating systems.
We argue that, counter-intuitively, the most security-conscious approach to OS
security is for vendors to ignore the expected security impact of updates.  In
other words, security updates should not be regarded as a special kind of
bug-fix update; instead, in core OS software, security bugs and normal bugs
should be treated as indistinguishable for most practical purposes.

Instead of focusing on security updates, we argue that OS vendors and security
groups should seek update technologies that make it feasible to distribute
updates for all disclosed OS bugs in a timely manner.  For example, \emph{hot
update} technology allows a running software system, such as an OS kernel, to
be updated with a minimal amount of disruption.

The rest of this paper is organized as follows: The next section describes two
notable historical events and what they can teach us about security impact
predictions.  Section~\ref{evaluation} presents evidence that depending on
security impact predictions is risky.  Section~\ref{hot-updates} discusses how
hot updates present a superior alternative to focusing on security impact
predictions.  Section~\ref{related-work} discusses related work, and
Section~\ref{conclusion} concludes.

\section{Lessons from history}

\subsection{Exploits can require extremely little}

The UNIX program \texttt{sudo} allows specified users to run specified
commands with elevated privileges, according to a security policy
defined in advance by the system administrator.  On February 19, 2001,
version 1.6.3p6 of \texttt{sudo} was released, correcting a bug in the
program's \texttt{do\_syslog} function that could cause it to perform
an out-of-bounds read operation and thus crash with a segmentation
fault.

The bug causes \texttt{sudo} to send inappropriate areas of the heap
to the UNIX system log function, \texttt{syslog}.  This bug leads only
to out-of-bounds read operations from memory, apart from a single NUL
byte written to memory before each call to \texttt{syslog}---and
immediately thereafter restored to the byte's previous value.  The
crash occurs on a read operation when this process reaches the end of
the heap.

The narrow reach of this bug led many people to conclude that it did
not threaten security.  Surely if ever there were a bug that could not
be exploited, a bug that replaces a single byte with NUL, only to
immediately restore it, would be a leading candidate.  Security expert
Florian Weimer called an exploit ``highly unlikely'' after a detailed
analysis of the bug~\cite{sudo-dismissal}.

Nevertheless, the bug can be exploited to achieve arbitrary execution.  In
November 2001, an exploit became public~\cite{vudo}.  With a thorough
understanding of the internal operation of \texttt{malloc} memory allocation,
even this most narrowly circumscribed bug can be successfully exploited to gain
full administrator privileges.

This case exemplifies the difficulty of accurately dismissing any bug in core
OS software, even a seemingly mild one, as not posing a security issue.

\subsection{Bad impact predictions cause problems}

Debian is one of the oldest and largest Linux distributions; it was started in
August 1993, and leading estimates indicate that roughly
35\%~\cite{linux-counter} of Linux machines run Debian or one of the
distributions built on top of Debian, such as Ubuntu.  The Debian project has
had two security compromises of its server infrastructure in its 15-year
history, on November 19, 2003~\cite{debian-nov-2003} and July 12,
2006~\cite{debian-jul-2006}.  The 2003 incident was only possible because of a
false reliance on security impact predictions.

The 2003 compromise took advantage of a bug for which a patch was available on
September 24, eight weeks before the attack~\cite{debian-analysis}.  The bug
had not been widely fixed by the time of the attack because no one knew that it
could be exploited; it was not classified as a security bug.

The attack was only discovered because the attackers left behind unusual log
messages, which were noticed by a Debian system
administrator~\cite{discovered}.  The Debian security team was then able to
shut down the machines, locate the exploit code, disassemble it, and identify
the nature of the exploit.  Had the attackers' rootkit been more subtle, had
they removed the exploit code before the machines were shut down, or had Debian
not possessed the expertise required to disassemble and reverse-engineer the
exploit, the attackers could have gone on to the next compromise without having
alerted anyone to the bug's security implications.  The attackers quite likely
did so with softer targets during the previous eight weeks.

If the Debian system administrators had applied all Linux kernel bug fixes
promptly, instead of only the bug fixes known to have security implications,
then the attackers would have failed.  With hot update technology, described in
Section~\ref{hot-updates}, this update policy can potentially be practiced with
minimal disruption.

\section{Evaluation}
\label{evaluation}

We examine two aspects of contemporary security impact information.  We use the
Linux kernel for this evaluation due to its transparent development processes
and widespread use.

In Section~\ref{delays}, we look at how commonly bugs are discovered
to have security implications long after the bug and its corresponding
patch have been publicly disclosed.

In Section~\ref{completeness}, we look at how commonly the security
consequences of bugs are never known to those individuals and organizations who
track security impact information.  In particular, we look at whether it is
advisable to treat the leading security vulnerability list, the Common
Vulnerabilities and Exposures (CVE)~\cite{CVE} list, as a complete list of the
disclosed bugs that have severe security implications.

\subsection{Delays before true impact is known}
\label{delays}

The purpose of this study was to investigate how often
initially-inaccurate security impact information results in bugs with
security consequences being overlooked.

\subsubsection{Methodology}

We define the \emph{impact delay} of a bug to be the period of time
between when the bug was disclosed (in this study, via a Linux patch)
and when its security implications were identified (i.e., the bug was
assigned a CVE number).

In this study, we identified instances of Linux kernel vulnerabilities
with large impact delay.  We generated our impact delay data using the
following process:

First, we created a list of all of the Linux kernel vulnerabilities added to
the Common Vulnerabilities and Exposures list during a three year period, from
January 2006 to December 2008.  We then found the Linux kernel patch
corresponding to each of these vulnerabilities and looked at the date that each
patch was finalized for inclusion into the Linux kernel\footnote{Specifically,
we used the date that the patch was added to either the mainline Linux
kernel~\cite{Linus} or one of the {\tt -stable}~\cite{stable} branches.}.

By comparing the date of the bug patch with the date that the bug CVE was
assigned, we found bugs whose security consequences were not recognized until
many weeks after the bugs were initially disclosed.

Requesting a CVE number for a new vulnerability normally takes less
than one business day, but we ignore vulnerabilities with less than
two weeks of impact delay in our analysis, in order to be
conservative.

\subsubsection{Results}

\begin{figure}
\begin{center}
\includegraphics[scale=0.40]{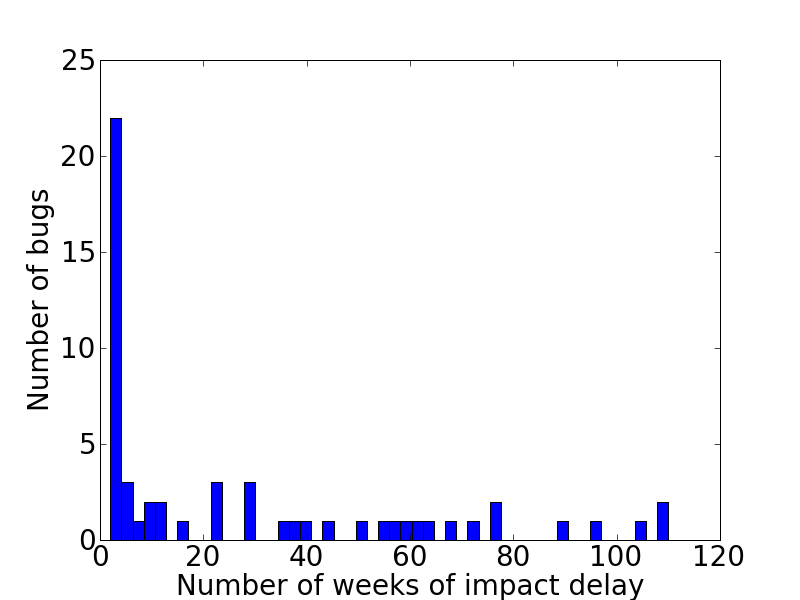}
\caption{Number of bugs discovered to be security bugs long after bug disclosure, from January 2006 to December 2008\label{fig:histograph}}
\end{center}
\end{figure}

Of the 218 Linux kernel CVEs from the studied interval, 25.7\% (56)
had more than two weeks of impact delay.  17.4\% (38) of the CVEs had
more than four weeks of impact delay, and 14.2\% (31) had more than
eight weeks of impact delay.  See Figure~\ref{fig:histograph} for the
distribution of CVEs with more than two weeks of impact delay.  The
raw data is available online~\cite{raw-data}.

These results indicate that many Linux bugs that pose a security risk
are only denoted as having security impact several weeks after the
bugs have been publicly disclosed.

To demonstrate that OS vendors commonly delay fixing bugs not
identified as having security impact, we studied the response of a
leading vendor, Red Hat, to the 30 bugs with the longest impact
delays, eight weeks or more.  Of these 30 bugs, 24 affected kernels
distributed by Red Hat.  We confirmed that none of these 24 bugs were
fixed by Red Hat until after their security consequences had been
discovered.

We also considered how many bugs, at any given time, had \emph{hidden
impact}---that is, had been disclosed as of that time, had no known
security impact at that time, but were found to have security impact
sometime before the end of 2008.

\begin{figure}
\begin{center}
\includegraphics[scale=0.40]{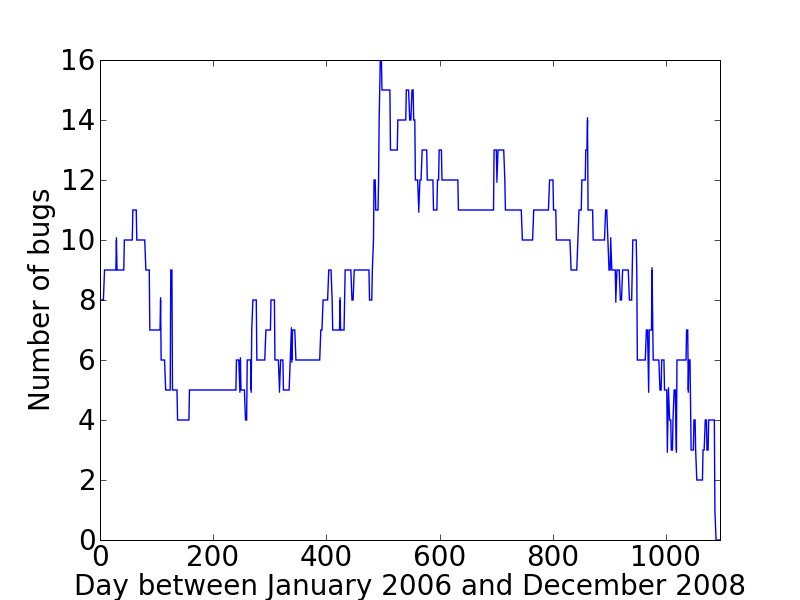}
\caption{The number of bugs with hidden impact on each day between
  January 2006 and December 2008}\label{fig:vuln-window}
\end{center}
\end{figure}

On each day in 2006, there were between 4 and 11 bugs with hidden
impact.  On each day in 2007, there were between 6 and 16 bugs with
hidden impact.  See Figure~\ref{fig:vuln-window}.  Note that, by our
definition of hidden impact, the number of bugs with hidden impact
must go to zero by the end of 2008; if we had CVE data for 2009, this
strong downward trend would presumably not occur.

Together these results show that bug disclosures, as commonly found in
the form of OS bug-fix updates, provide a rich vein of vulnerabilities
not publicly identified as such and consequently not widely patched.
In the next subsection, we explore how difficult it would be for an
attacker to tap this vein.

\subsection{Completeness of vulnerability lists}
\label{completeness}

The purpose of this study was to investigate how easy it is to find serious
security bugs which have been disclosed, but not fixed even on
``fully-updated'' end-user machines, because of incorrect security impact
predictions.

\subsubsection{Methodology}

We reviewed bug-fix patches affecting Linux kernel version 2.6.24.  We selected
this version simply because it was the first Linux kernel release of 2008.

We looked at patches with no known security consequences to determine whether
any of them actually have severe security consequences---in particular, whether
any of them enable an attacker to achieve arbitrary code execution with
administrator privileges.

\subsubsection{Results}

Within a few hours of review of the bug-fix patches affecting Linux kernel
version 2.6.24, we identified a commit from February 2008 with serious security
consequences (Git ID {\tt 7e3c396}, commit subject ``{\tt
sys\_remap\_file\_pages: fix ->vm\_file accounting}'').  At the time that we
conducted this review, this bug and its corresponding patch had been disclosed
for more than 10 months, yet it had no associated CVE number or record of any
security consequences.

We developed a privilege escalation exploit for this bug in a few hours; doing
so did not require any innovative techniques or extensive expertise.  The
exploit allows any user on a vulnerable system to gain full administrator
privileges on the system.

Since vendors use security impact predictions to determine which bug-fix
patches to distribute, the patch for this bug was not widely distributed, even
though other bug-fix patches from the same period were widely deployed.  Fedora
7, for example, is affected by this bug but never received an update for it,
which means that all Fedora 7 systems remained vulnerable to this exploit
through Fedora 7's end-of-life in June 2008.

We reported the security consequences of this bug in January 2009, and it was
assigned CVE-2009-0024 at that time.

We studied nearly year-old bug fixes to make our task more difficult;
as Figure~\ref{fig:histograph} shows, many more bugs have impact
delays of two weeks or four weeks than ten months.  Yet even on bugs
where no vulnerability had been identified nearly a year after
disclosure, we succeeded with little effort in identifying and
exploiting a vulnerability.  An attacker seeking to exploit
unidentified vulnerabilities in Linux bug-fix disclosures would have,
as Figure~\ref{fig:vuln-window} shows, between 4 and 16 bugs with
hidden impact waiting for him or her at any time in the last three years.

\section{Implications: Hot updates}
\label{hot-updates}

In this paper, we argue that OS vendors should not attempt to treat
security updates differently from other bug fix updates.
Unfortunately, distributing all updates with equal priority increases
the quantity of updates that system administrators are expected to
apply in a timely manner.

Applying more OS updates is problematic because of a long-standing problem with
how software updates are typically performed: currently, a program must be
restarted in order for it to be updated, which is disruptive.  This problem is
particularly severe for core OS software, such as the kernel itself, that
cannot normally be updated without rebooting the operating system.

Frequent OS reboots are costly since, in addition to any service availability
concerns, many system administrators want to monitor their systems during the
disruptive reboot process, in order to deal with any complications that arise.

Hot update techniques~\cite{OPUS, Ksplice, K42b, LUCOS, DynAMOS,
Ginseng} make it possible to correct bugs in a running program without
restarting the program or interfering with its operation.  The Ksplice
hot update system~\cite{Ksplice} has recently shown that it is
possible to transform many historical security patches into hot
updates with little or no programmer involvement.

If this progress can be extended, a hot update system could
potentially generate hot updates for all core OS bug-fix patches with
little programmer involvement.  Achieving this goal would make it
possible to stop relying on security impact predictions, which would,
as we have argued, improve security.

\section{Related Work}
\label{related-work}

Security researchers have surveyed known vulnerabilities, computing statistics
involving various dates, such as dates of first disclosure and of exploit
availability.  Rescorla~\cite{vuln-pool} analyzed vulnerability disclosure
rates to suggest that popular software contains many more vulnerabilities than
have been discovered so far.

Frei et al.~\cite{large-scale} found that about 90\% of vulnerabilities have
exploits available within days after disclosure, while fewer than 20\% have
exploits available before disclosure.

These results are consistent with our argument that hot update technology---or
more generally, the ability to apply updates for newly-discovered bugs
promptly---is important for improving security.

Like Linux vendors, Microsoft's Windows Update service~\cite{WindowsUpdate}
classifies updates into categories based on the perceived impact of the
updates, in order to encourage end-users and system administrators to install
high-priority updates more rapidly than low-priority updates.

\section{Conclusions}
\label{conclusion}

We have shown that, following the disclosure of many core OS bugs, weeks or
months lapse before they are identified as security bugs.  Based on historical
lessons and our own exploit investigation, we conclude that disclosed bugs
present a significant security risk until they are fixed with an update,
regardless of their perceived security impact.

Treating some disclosed bugs as being the only bugs with high security impact,
without conclusive proof, weakens OS security by engendering a false sense of
security while providing attackers with the information and time that they need
to compromise systems.

Research into improved update technology, such as hot updates, has the
potential to eliminate reliance on security impact predictions, which would be
a notable security improvement.

\section*{Acknowledgments}

We thank Frans Kaashoek for helpful comments.

\small{

}
\end{document}